\documentclass[a4paper]{PoS}
\pdfoutput=1

\usepackage{amssymb,amsmath,fontenc,times,mathptmx,graphicx}
\usepackage{multirow}
\usepackage{array,booktabs,colortbl,multirow}
\usepackage{colortbl,xcolor,color}
\usepackage{rotating}
\usepackage{placeins}
\usepackage{wasysym}
\usepackage{floatrow}

\newfloatcommand{capbtabbox}{table}[][\FBwidth]

\newcommand{\cmrule}{\midrule[0.25mm]}
\newcommand{\lrD}{~\!\overset{\leftrightarrow}{\hspace{-0.1cm}D}\!}

\title{Electroweak precision constraints at present and future colliders}

\ShortTitle{Electroweak precision constraints at present and future colliders}

\author{\speaker{Jorge de Blas}\\
      INFN, Sezione di Roma\\
      E-mail: \email{jorge.deblasmateo@roma1.infn.it}}

\author{Marco Ciuchini\\
        INFN, Sezione di Roma Tre\\ 
        E-mail: \email{marco.ciuchini@roma3.infn.it}
        }
      
\author{Enrico Franco\\
        INFN, Sezione di Roma\\ 
        E-mail: \email{enrico.franco@roma1.infn.it}}
        
\author{Satoshi Mishima\\
        Theory Center, IPNS, KEK\\
        E-mail: \email{satoshi.mishima@kek.jp}}
        
\author{Maurizio Pierini\\
        CERN\\
        E-mail: \email{maurizio.pierini@cern.ch}}
        
\author{Laura Reina\\
        Physics Department, Florida State University\\
        E-mail: \email{reina@hep.fsu.edu}}
        
\author{Luca Silvestrini\\
        INFN, Sezione di Roma\\ 
        E-mail: \email{luca.silvestrini@roma1.infn.it}}

\abstract{We revisit the global fit to electroweak precision observables in the Standard Model
and present model-independent bounds on several general new physics scenarios. 
We present a projection of the fit based on the expected experimental improvements at future 
$e^+ e^-$ colliders, and compare the constraining power of some of the different experiments 
that have been proposed. All results have been obtained with the {\tt HEPfit} code.
}

\FullConference{38th International Conference on High Energy Physics\\
		  3-10 August 2016\\
		 Chicago, USA}
		 
\begin{document}

\section{Introduction}
\label{sec:intro}

With the discovery of a Standard Model (SM) -like Higgs boson in 2012 at the Large Hadron Collider (LHC)
all the input parameters of the SM have been experimentally measured. Thus, the SM prediction for any
given observable is now unambiguous, and the same applies to limits on the size of possible
new physics (NP) contributions. 
In particular, the good agreement between the SM and the very precise measurements of the electroweak precision observables (EWPO) imposes strong constraints on NP modifying the electroweak sector. 
In these proceedings we review the current status of the electroweak precision data (EWPD) constraints on physics beyond the SM, 
and study the improvements expected at future $e^+ e^-$ colliders.
We adopt a model-independent approach and present our results in terms of different 
parameterizations of NP: oblique $S$, $T$, $U$ parameters, NP in the $Z\bar{b}b$ interactions, modified
Higgs couplings to vector bosons, and the more general case of the dimension 6 SM effective field theory. In the next 
section we summarize the status of the SM fit to EWPD and introduce the different future $e^+ e^-$ colliders
that have been proposed and we use in our analyses. In Section~\ref{sec:Constr_pres_fut} we present the projected sensitivities to the NP scenarios mentioned above at such facilities, comparing them with current constraints.
We close the paper with a short summary and conclusions.


\section{Electroweak precision observables at present and future colliders}
\label{sec:EWPO_present_future}

The bulk of the EWPO comprises the $Z$-pole measurements taken at LEP and SLD, and the $W$-boson properties measured at LEP2 and Tevatron. 
The fit to EWPD also receives inputs from the LHC via the determinations of the Higgs-boson and top-quark masses (the latter also measured at the Tevatron), as well as from experiments measuring the running of the electromagnetic constant, parameterized in terms of the 5-flavour hadronic contribution, $\Delta \alpha_{\mathrm{had}}^{(5)}(M_Z)$, and the strong coupling constant, $\alpha_S(M_Z)$. In Ref.~\cite{deBlas:2016ojx} we presented the most up-to date fit of the SM to current EWPD, using the {\tt HEPfit} code. The results are summarized in Table~\ref{tab:SMfit}, where we see how the indirect determinations of the SM input parameters from the fit
are consistent with the experimental observations at the $\sim 1~\!\sigma$ level. Also, with the exception of the $-2.6~\!\sigma$ discrepancy on the forward-backward asymmetry of the $b$ quark, $A_{\rm FB}^{0,b}$, all the SM predictions for EWPO agree with the data at $\lesssim 2~\!\sigma$. It is also noteworthy that, using current theoretical calculations, the overall (theoretical+parametric) uncertainty of the predictions is well below the experimental errors.

\begin{table}[t]
{\footnotesize
\begin{center}
\begin{tabular}{lcccc}
\toprule
& Measurement & Posterior & Prediction &Pull \\
\cmrule
$\alpha_s(M_Z)$ & $ 0.1179 \pm 0.0012 $ & $ 0.1180 \pm 0.0011 $  & $ 0.1185 \pm 0.0028 $ & -0.2  \\ 
$\Delta\alpha_{\rm had}^{(5)}(M_Z)$ & $ 0.02750 \pm 0.00033 $ & $ 0.02747 \pm 0.00025 $  & $ 0.02743 \pm 0.00038 $ & 0.04  \\ 
$M_Z$ [GeV] & $ 91.1875 \pm 0.0021 $ & $ 91.1879 \pm 0.0020 $  & $ 91.199 \pm 0.011 $ & -1.0 \\ 
$m_t$ [GeV]  & $ 173.34 \pm 0.76 $ & $ 173.61 \pm 0.73 $  & $ 176.6 \pm 2.5 $ & -1.3 \\ 
$m_H$ [GeV]  & $ 125.09 \pm 0.24 $ & $ 125.09 \pm 0.24 $  & $ 102.8 \pm 26.3 $ & 0.8  \\
\cmrule
$M_W$ [GeV]  & $ 80.385 \pm 0.015 $ & $ 80.3644 \pm 0.0061 $  & $ 80.3604 \pm 0.0066 $ & 1.5  \\ 
$\Gamma_{W}$ [GeV]  & $ 2.085 \pm 0.042 $ & $2.08872 \pm 0.00064 $  & $ 2.08873 \pm 0.00064 $ & -0.2  \\ 
$\sin^2\theta_{\rm eff}^{\rm lept}(Q_{\rm FB}^{\rm had})$ & $ 0.2324 \pm 0.0012 $ & $ 0.231464 \pm 0.000087 $  & $ 0.231435 \pm 0.000090 $ & 0.8  \\
$P_{\tau}^{\rm pol}={A}_\ell$ & $ 0.1465 \pm 0.0033 $ & $ 0.14748 \pm 0.00068 $  & $ 0.14752 \pm 0.00069 $ & -0.4  \\ 
$\Gamma_{Z}$ [GeV] & $ 2.4952 \pm 0.0023 $ & $ 2.49420 \pm 0.00063 $  & $ 2.49405 \pm 0.00068 $ & 0.5 \\ 
$\sigma_{h}^{0}$ [nb] & $ 41.540 \pm 0.037 $ & $ 41.4903 \pm 0.0058 $  & $ 41.4912 \pm 0.0062 $ & 1.3 \\
$R^{0}_{\ell}$ & $ 20.767 \pm 0.025 $ & $ 20.7485 \pm 0.0070 $  & $ 20.7472 \pm 0.0076 $ & 0.8 \\ 
$A_{\rm FB}^{0, \ell}$  & $ 0.0171 \pm 0.0010 $ & $ 0.01631 \pm 0.00015 $  & $ 0.01628 \pm 0.00015 $ & 0.8 \\ 
${A}_{\ell}$ (SLD) & $ 0.1513 \pm 0.0021 $ & $ 0.14748 \pm 0.00068 $  & $ 0.14765 \pm 0.00076 $ & 1.7 \\ 
${A}_c$ & $ 0.670 \pm 0.027 $ & $ 0.66810 \pm 0.00030 $  & $ 0.66817 \pm 0.00033 $ & 0.02 \\ 
${A}_b$ & $ 0.923 \pm 0.020 $ & $ 0.934650 \pm 0.000058 $  & $ 0.934663 \pm 0.000064 $ & -0.6 \\ 
$A_{\rm FB}^{0, c}$ & $ 0.0707 \pm 0.0035 $ & $ 0.07390 \pm 0.00037 $  & $ 0.07399 \pm 0.00042 $ & -0.9 \\ 
$A_{\rm FB}^{0, b}$ & $ 0.0992 \pm 0.0016 $ & $ 0.10338 \pm 0.00048 $  & $ 0.10350 \pm 0.00054 $ & -2.6 \\ 
$R^{0}_{c}$ & $ 0.1721 \pm 0.0030 $ & $ 0.172228 \pm 0.000023 $  & $ 0.172229 \pm 0.000023 $ & -0.05 \\ 
$R^{0}_{b}$ & $ 0.21629 \pm 0.00066 $ & $ 0.215790 \pm 0.000028 $  & $ 0.215788 \pm 0.000028 $ & 0.7 \\
\bottomrule
\end{tabular}
\end{center}
}
\vspace{-0.4cm}
\caption{
  Experimental measurement, posterior, prediction, and pull for the 5 input
  parameters ($\alpha_s(M_Z)$, $\Delta \alpha^{(5)}_{\mathrm{had}}(M_Z)$, $M_Z$,
  $m_t$, $m_H$), and for the main EWPO considered in the SM fit. The values in
  the column \emph{Prediction} are determined without using the
  experimental information for the corresponding observable.} 
\label{tab:SMfit}
\end{table}

Several future $e^+ e^-$ colliders have been proposed to improve the precision of the electroweak (and Higgs-boson) observables. In this study we consider the FCCee project at CERN~\cite{Gomez-Ceballos:2013zzn},  the ILC in Japan~\cite{Fujii:2015jha}\footnote{While a $\sqrt{s}\approx91$ GeV run with optimal luminosity would require a machine upgrade from the current Technical Design Report, we still keep the ILC in our comparisons but include only the improvements in non $Z$-pole observables.}, and the CEPC in China~\cite{CEPC}. For completeness, we also include in the study the expected improvements in the top and $W$ masses at the high luminosity LHC (HL-LHC).
Some of these machines would improve the precision of many EWPO by one order of magnitude ---see \cite{deBlas:2016ojx} for details--- being able to test the SM predictions beyond the 2-loop level. 
To optimize the sensitivity to NP at such facilities it will be therefore necessary to compute the general 3-loop (maybe even the leading 4-loop) contributions to SM predictions, in order to match the experimental precision. In the next section we discuss the projected sensitivities to physics beyond the SM by comparing the statistical uncertainties for the corresponding NP parameters from the fit. We assume the future experiments will measure the SM predictions, with the errors reported in Table 19 in Ref.~\cite{deBlas:2016ojx}. We also use the expected improvement in the determination of $\Delta \alpha_{\mathrm{had}}^{(5)}(M_Z)$ and $\alpha_S(M_Z)$, with projected uncertainties of $\pm5\times10^{-5}$ and $\pm 0.0002$, respectively, to reduce the parametric uncertainties in the SM predictions. Finally, we present the results assuming the future SM theory errors (see Table 21 in Ref.~\cite{deBlas:2016ojx}), and study the impact of such uncertainties by comparing with a scenario where we assume that the SM errors are subdominant.


\section{Electroweak constraints on new physics: present and future}
\label{sec:Constr_pres_fut}

\begin{figure}[t]
\begin{center}
  \begin{tabular}{c c c}
 \hspace{-0.8cm}\includegraphics[width=.33\textwidth]{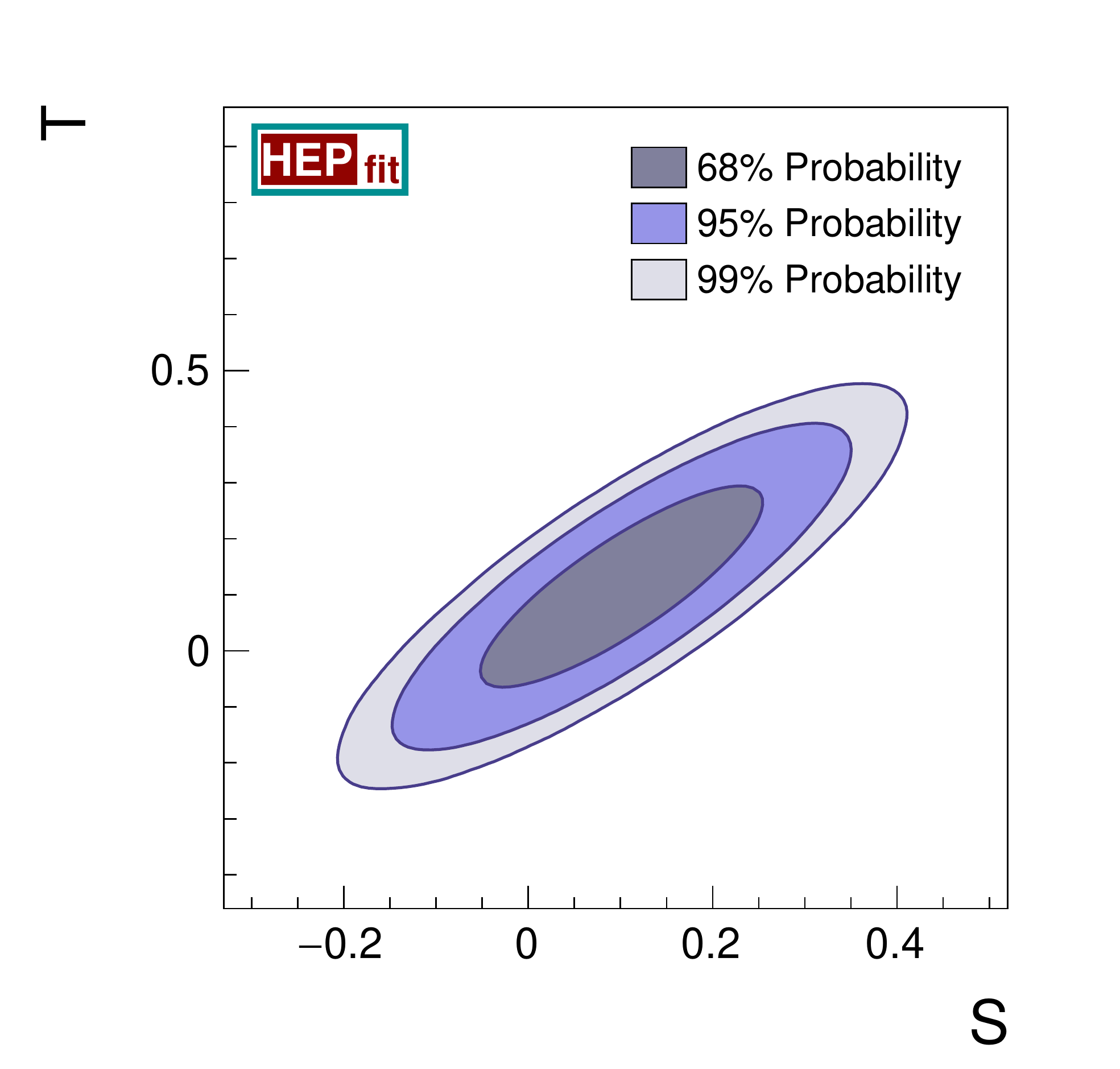} 
&
  \hspace{-0.6cm}\includegraphics[width=.33\textwidth]{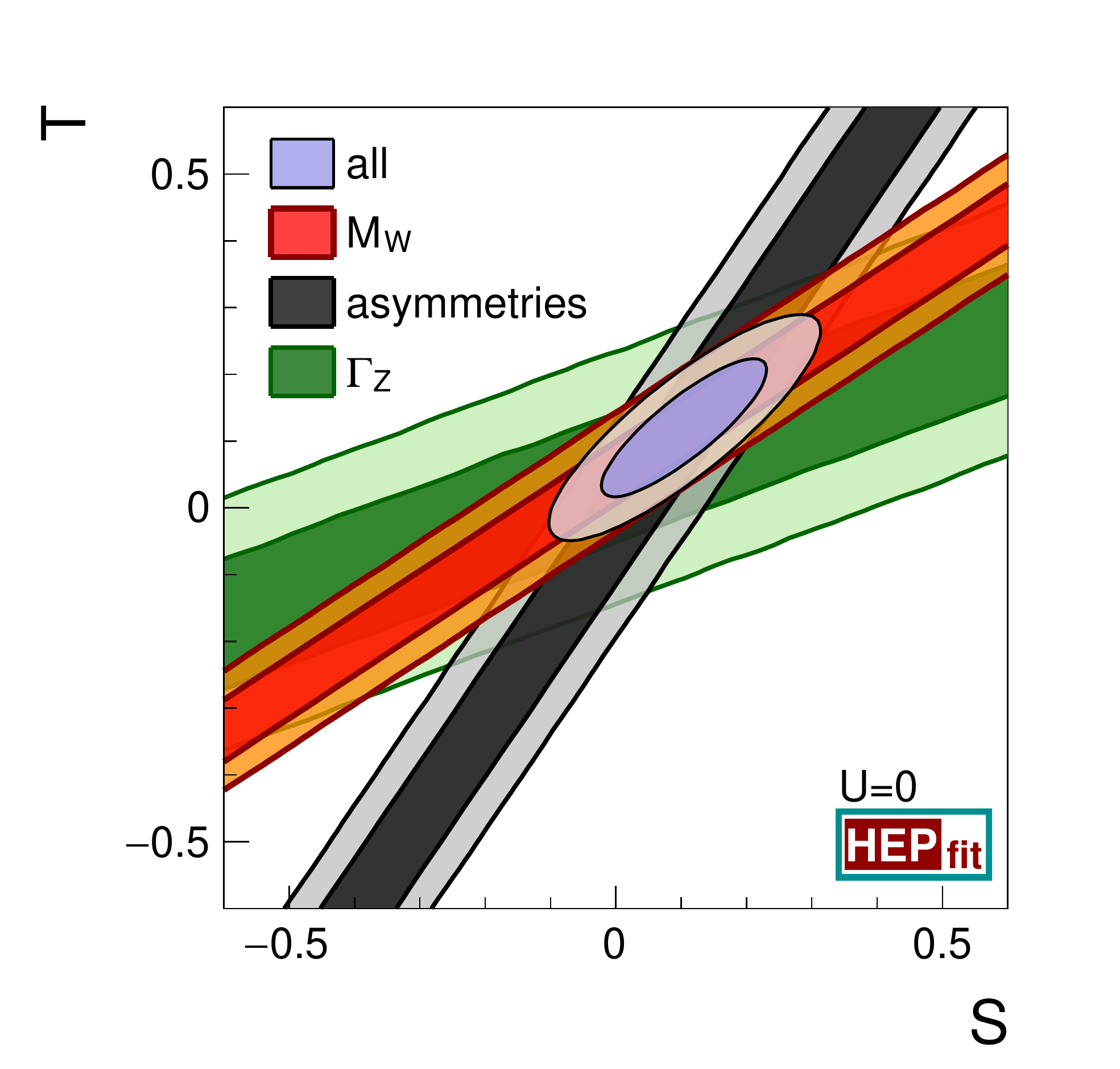} 
&
 \hspace{-0.8cm}\includegraphics[width=.422\textwidth]{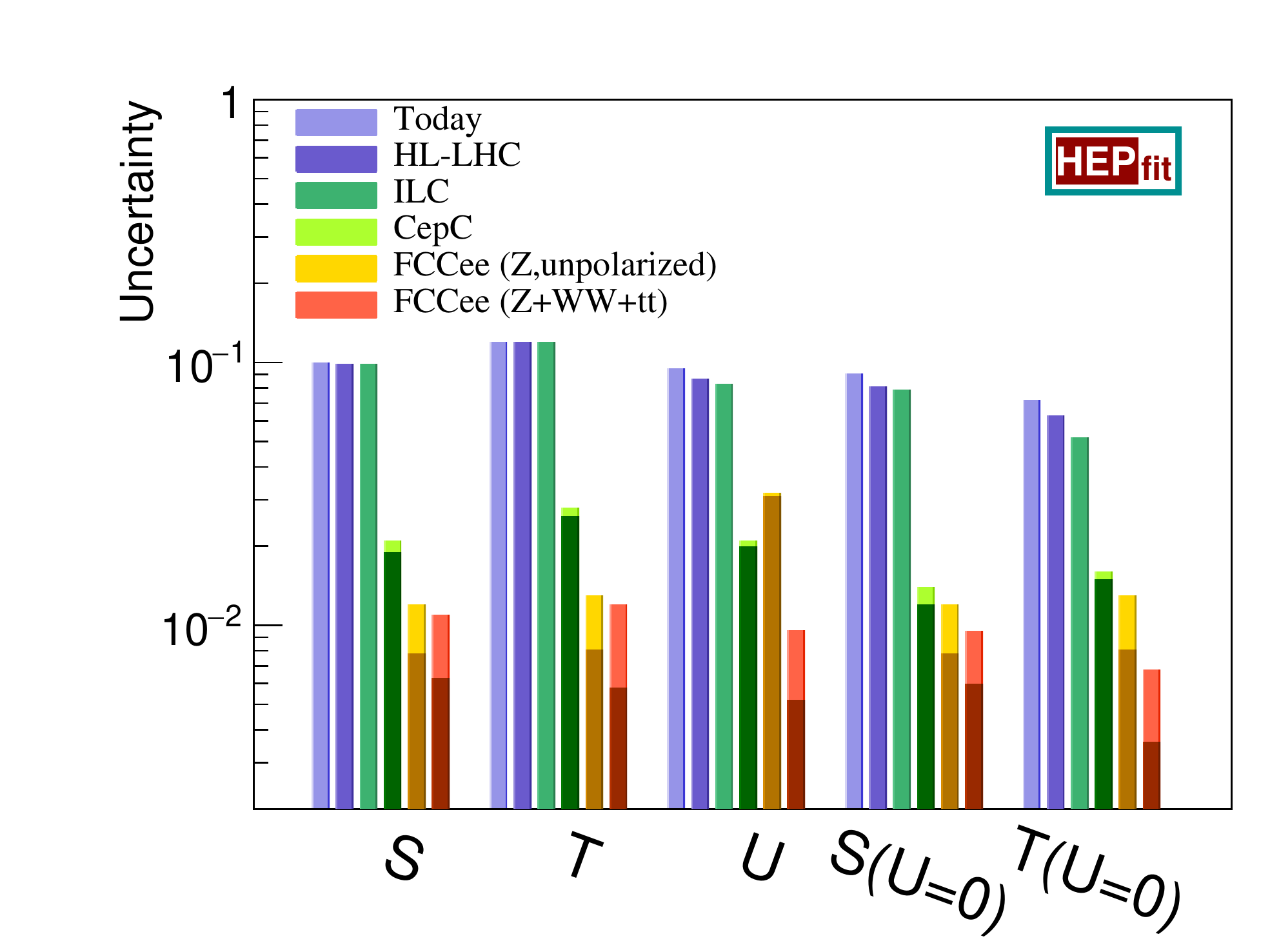}
  \end{tabular}
  \vspace{-0.65cm}
  \caption{(Left) $68\%$, $95\%$, and $99\%$ probability contours for the
    $S$ and $T$ parameters. (Center)
    $68\%$ and $95\%$ probability contours for $S$ and $T$
    fixing $U=0$, together with the individual constraints from $M_W$,
    the asymmetry parameters $\sin^2\theta_{\rm eff}^{\rm lept}$,
    $P_\tau^{\rm pol}$, $A_f$, and $A_{\rm FB}^{0,f}$ with $f=\ell,c,b$, 
    and $\Gamma_Z$.
    (Right) Expected sensitivities to $S,~T,~U$
    at future colliders. Different shades of the same colour correspond to
    results including or neglecting the future theoretical
    uncertainties. }
  \label{fig:Oblique}
  \end{center}
\end{figure}

We first study scenarios in which NP only modifies the electroweak gauge-boson propagators. 
From the point of view of the EWPO, such effects can be described in terms of 3
parameters, the well-known $S$, $T$, and $U$ oblique parameters~\cite{Peskin:1991sw}. 
In theories where the electroweak symmetry breaking is linearly realized, however, 
$U$ is expected to be much smaller than $S$ and $T$.
Hence in Tables~\ref{tab:STUfit} and \ref{tab:STfit}, and Fig.~\ref{fig:Oblique} we show the results for the $STU$ fit with and without fixing $U=0$. We also show in Fig.~\ref{fig:Oblique} the expected improvements in the sensitivity at future experiments. As can be seen, there is no evidence of NP in the current data. Future facilities like the FCCee and CepC would be sensitive to effects approximately one order of magnitude smaller than those that can be probed with current data. It is also apparent that, given the high precision expected at the FCCee, the future SM theoretical uncertainties would still be a limiting factor, reducing the sensitivity to $S,~T,~U$ in some cases by up to a factor of 2.
\begin{figure}[h]
\begin{floatrow}
\capbtabbox{
\centering
\begin{tabular}{c c rrr}
 \toprule
 & Result & \multicolumn{3}{c}{Correlation Matrix} \\ 
 \cmrule
$S$ & $ 0.09 \pm 0.10 $ & $1.00$ \\ 
$T$ & $ 0.10 \pm 0.12 $ & $0.86$ & $1.00$ \\ 
$U$ & $ 0.01 \pm 0.09 $ & $-0.54$ & $-0.81$ & $1.00$ \\
\bottomrule
 \end{tabular}
 }{
\caption{Results of the fit for the oblique parameters $S$, $T$, and $U$.}
\label{tab:STUfit}
}
\capbtabbox{
\centering
\begin{tabular}{c c rr}
 \toprule
 & Result & \multicolumn{2}{c}{Correlation Matrix} \\ 
 \cmrule
$S$ & $ 0.10 \pm 0.08 $ & $1.00$ \\ 
$T$ & $ 0.12 \pm 0.07 $ & $0.86$ & $1.00$ \\ 
\bottomrule
 \end{tabular}
 }{
\caption{Results of the fit for the oblique parameters $S$ and $T$, fixing $U=0$.}
\label{tab:STfit}
}
\end{floatrow}
\end{figure}

\begin{figure}[t]
\begin{center}
  \begin{tabular}{c c c}
 \hspace{-0.8cm}\includegraphics[width=.33\textwidth]{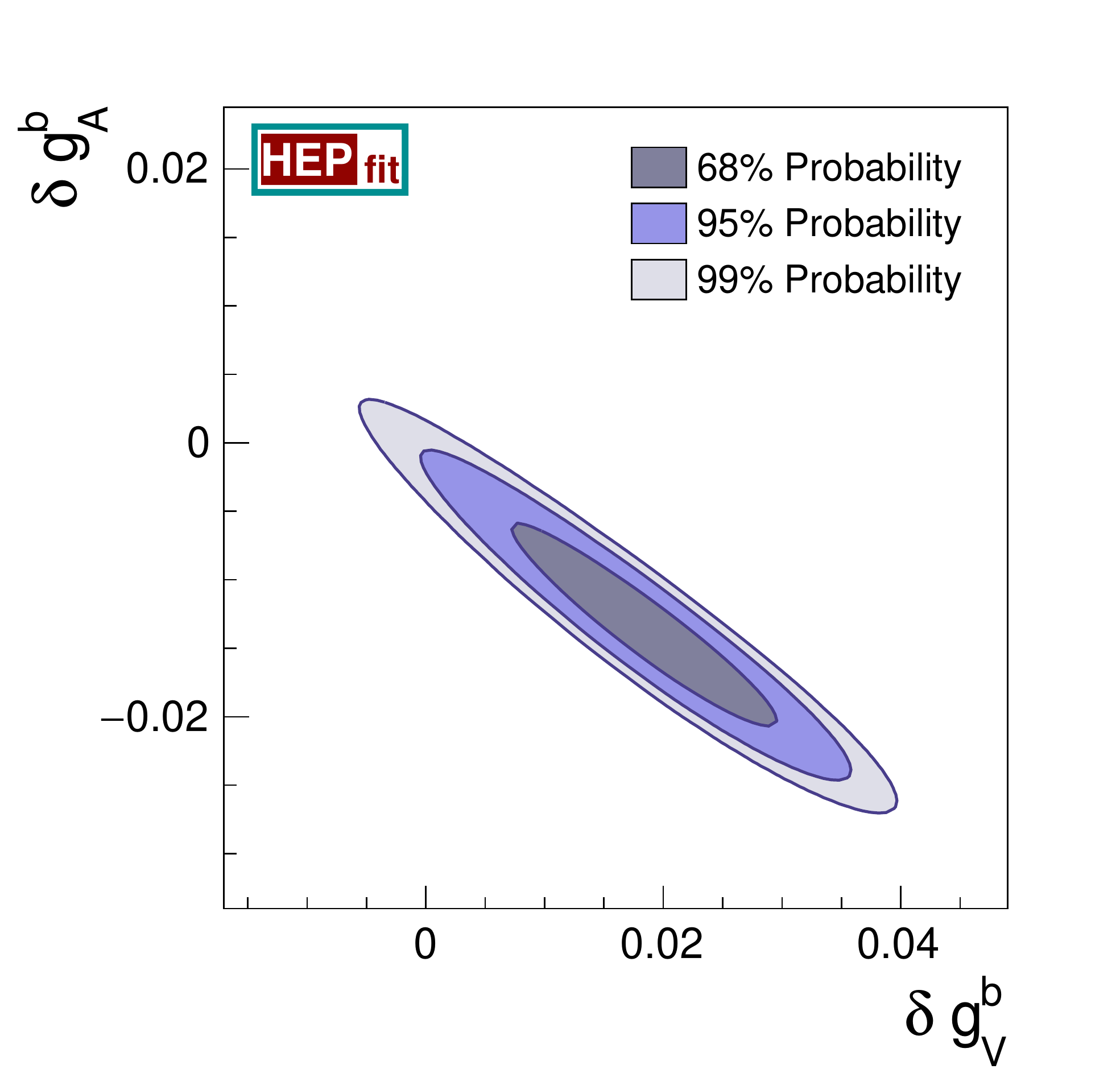} 
&
  \hspace{-0.3cm}\includegraphics[width=.33\textwidth]{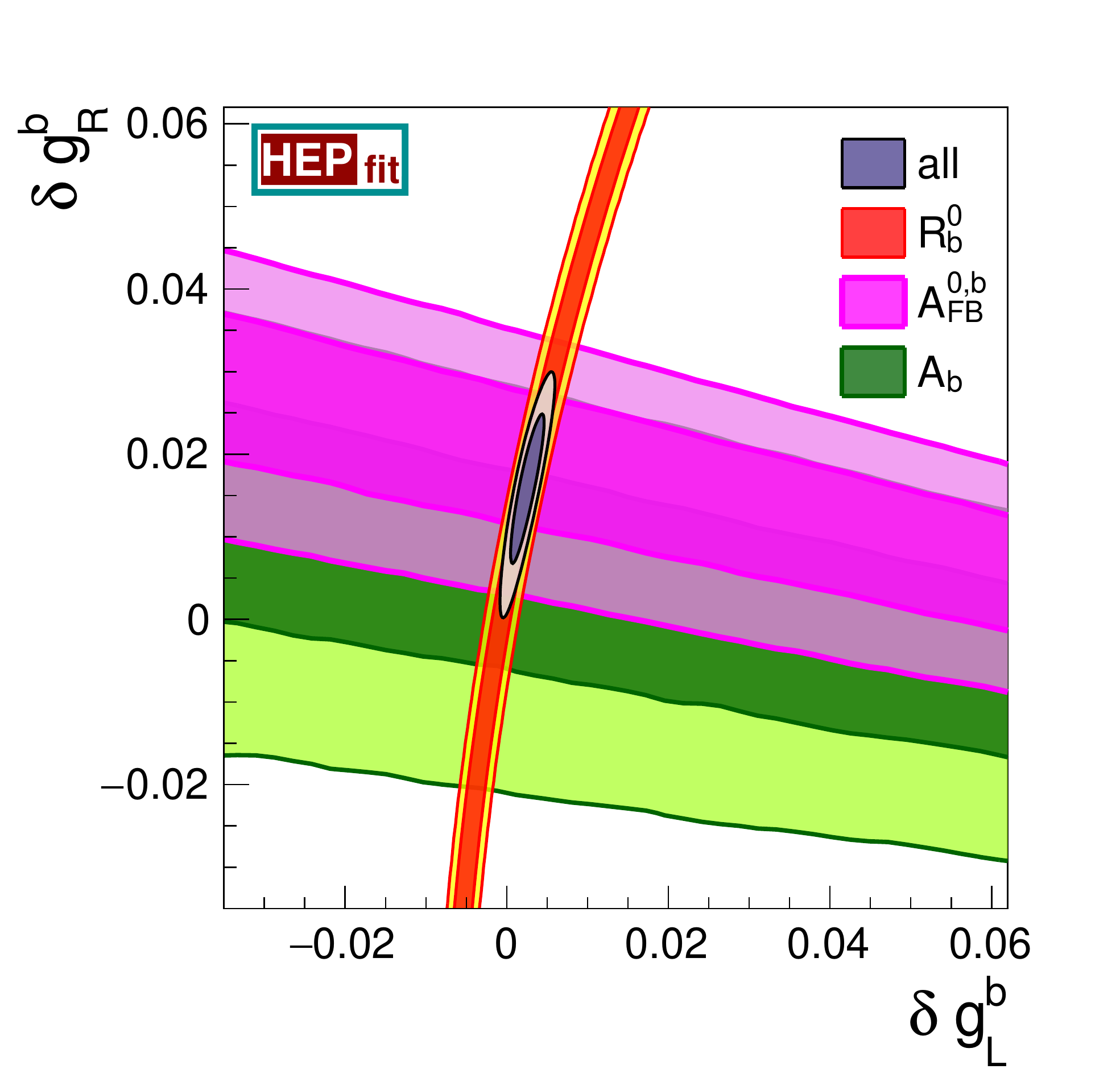} 
&
 \hspace{-0.3cm}\includegraphics[width=.344\textwidth]{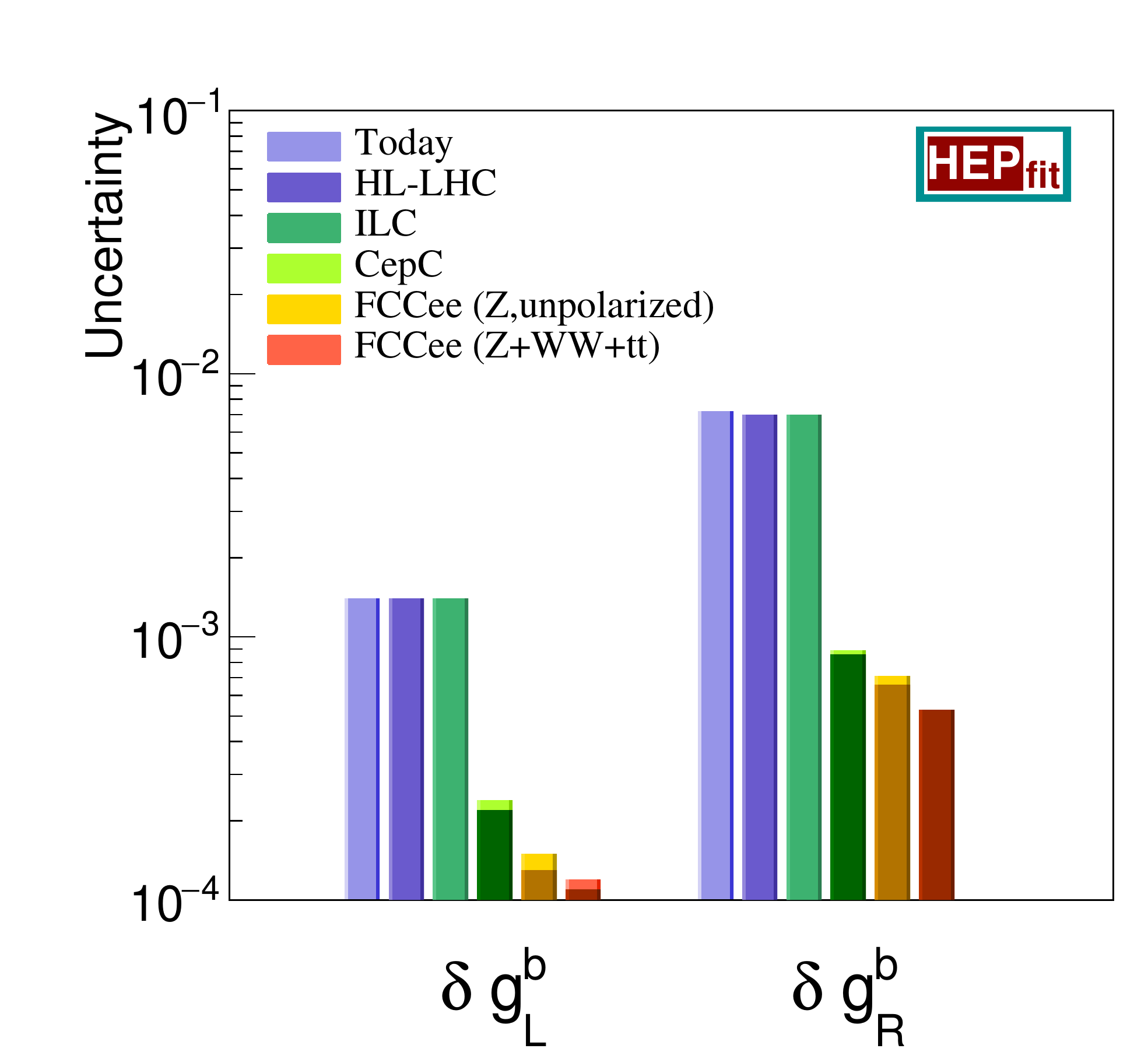}
  \end{tabular}
  \vspace{-0.65cm}
  \caption{(Left) $68\%$, $95\%$, and $99\%$ probability contours for the
    $\delta g_V^b$, $\delta g_A^b$ couplings. (Center)
    $68\%$ and $95\%$ probability contours for $\delta
    g_R^b$, $\delta g_L^b$, together with the constraints from $R_b^0$,
    $A_{FB}^0$ and $A_b$.
    (Right) Expected sensitivities to $\delta g_R^b$, $\delta g_L^b$
    at future colliders. Different shades of the same colour correspond to
    results including or neglecting the future theoretical
    uncertainties. }
  \label{fig:Zbb}
  \end{center}
\end{figure}

Motivated by the $-2.6~\sigma$ discrepancy in $A_{\rm FB}^{0,b}$, it is interesting to consider the possibility that the leading NP effects in EWPO manifest in extra contributions to the $Z\bar{b}b$ couplings,
\begin{equation}
g_a^b=g_a^{b~\mathrm{SM}} + \delta g_a^{b},~~a=L,R~\mathrm{or}~V,A.
\end{equation}
The results of the fit to EWPD provide four solutions for $\delta g_{a}^b$, but two of them are disfavored by the heavy flavour LEP2 data. The two surviving solutions are characterized by a relatively small $\delta g_{L}^b$, due to the $R_b$ constraints, and a sizable contribution to $\delta g_{R}^b$, needed to solve the $A_{\rm FB}^{0,b}$ anomaly. In Tables~\ref{tab:ZbbLR} and \ref{tab:ZbbVA} and Fig.~\ref{fig:Zbb} we show the results for the solution that is closer 
to the SM. While current data is barely consistent with the SM at $95\%$ probability, the order of magnitude 
improvement at the FCCee or CepC ---also shown in Fig.~\ref{fig:Zbb}--- would allow to confirm whether the $A_{\rm FB}^{0,b}$ is a probe of NP
or simply an outlier. 
\begin{figure}[h]
\begin{floatrow}
\capbtabbox{
\centering
\begin{tabular}{c c rr}
 \toprule
 & Result & \multicolumn{2}{c}{Correlation Matrix} \\ 
 \cmrule
$\delta g_{R}^{b}$ & $ 0.016 \pm 0.006 $ & $1.00$ \\ 
$\delta g_{L}^{b}$ & $ 0.002 \pm 0.001 $ & $0.90$ & $1.00$ \\ 
\bottomrule
 \end{tabular}
 }{
\caption{Results of the fit for the shifts in the left-handed and
right-handed $Zb\bar b$ couplings.}
\label{tab:ZbbLR}
}
\hspace{-0.5cm}\capbtabbox{
\centering
\begin{tabular}{c c rr}
 \toprule
 & Result & \multicolumn{2}{c}{Correlation Matrix} \\ 
 \cmrule
$\delta g_{V}^{b}$ & $ 0.018 \pm 0.007 $ & $1.00$ \\ 
$\delta g_{A}^{b}$ & $ -0.013 \pm 0.005 $ & $-0.98$ & $1.00$ \\ 
\bottomrule
 \end{tabular}
}{
\caption{Results of the fit for the shifts in the vector and axial-vector $Zb\bar b$ couplings.}
\label{tab:ZbbVA}
}
\end{floatrow}
\end{figure}

\begin{figure}[t]
\begin{center}
  \begin{tabular}{c c c}
 \hspace{-0.8cm}\includegraphics[width=.33\textwidth]{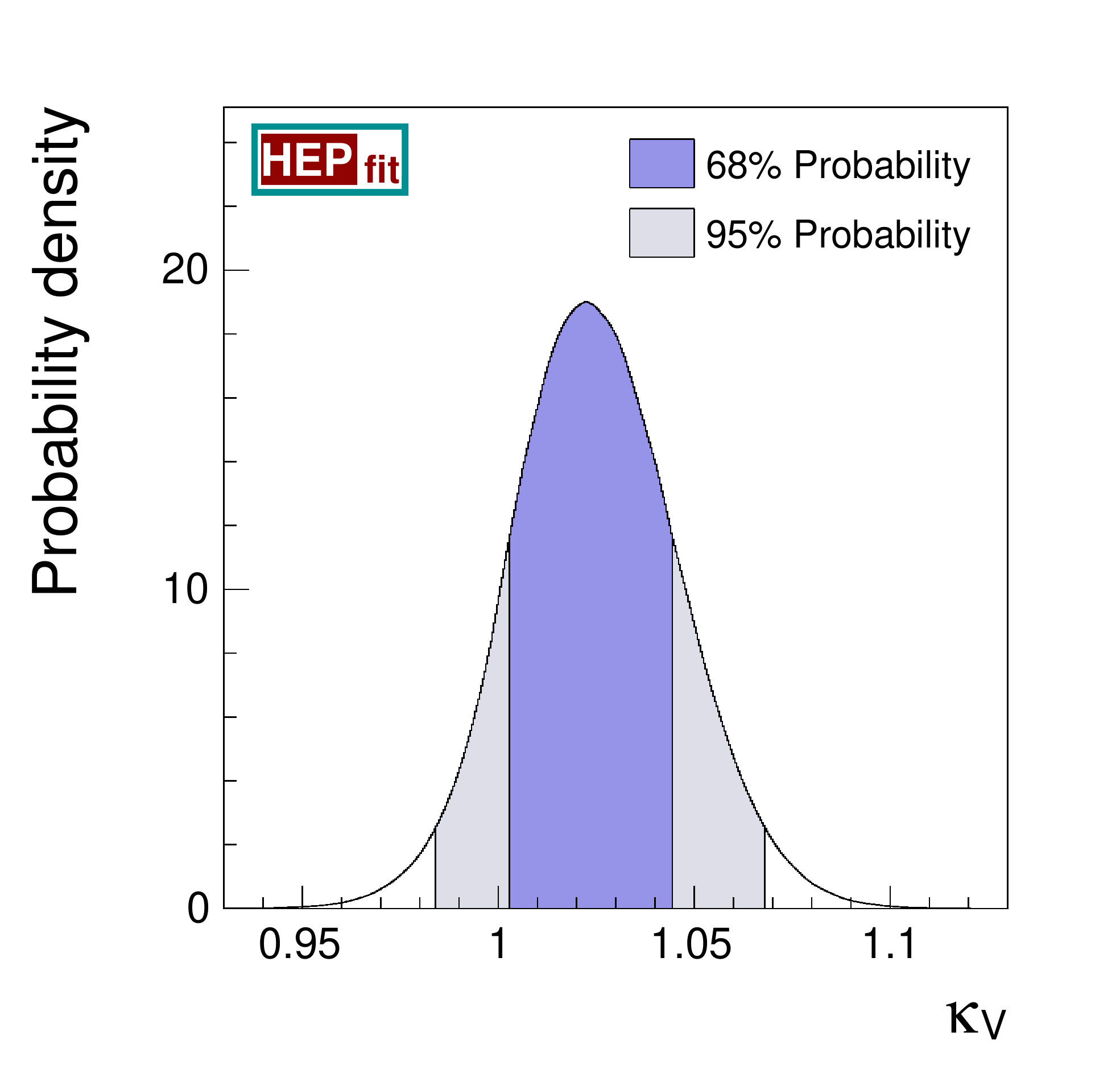} 
&
  \hspace{-0.3cm}\includegraphics[width=.33\textwidth]{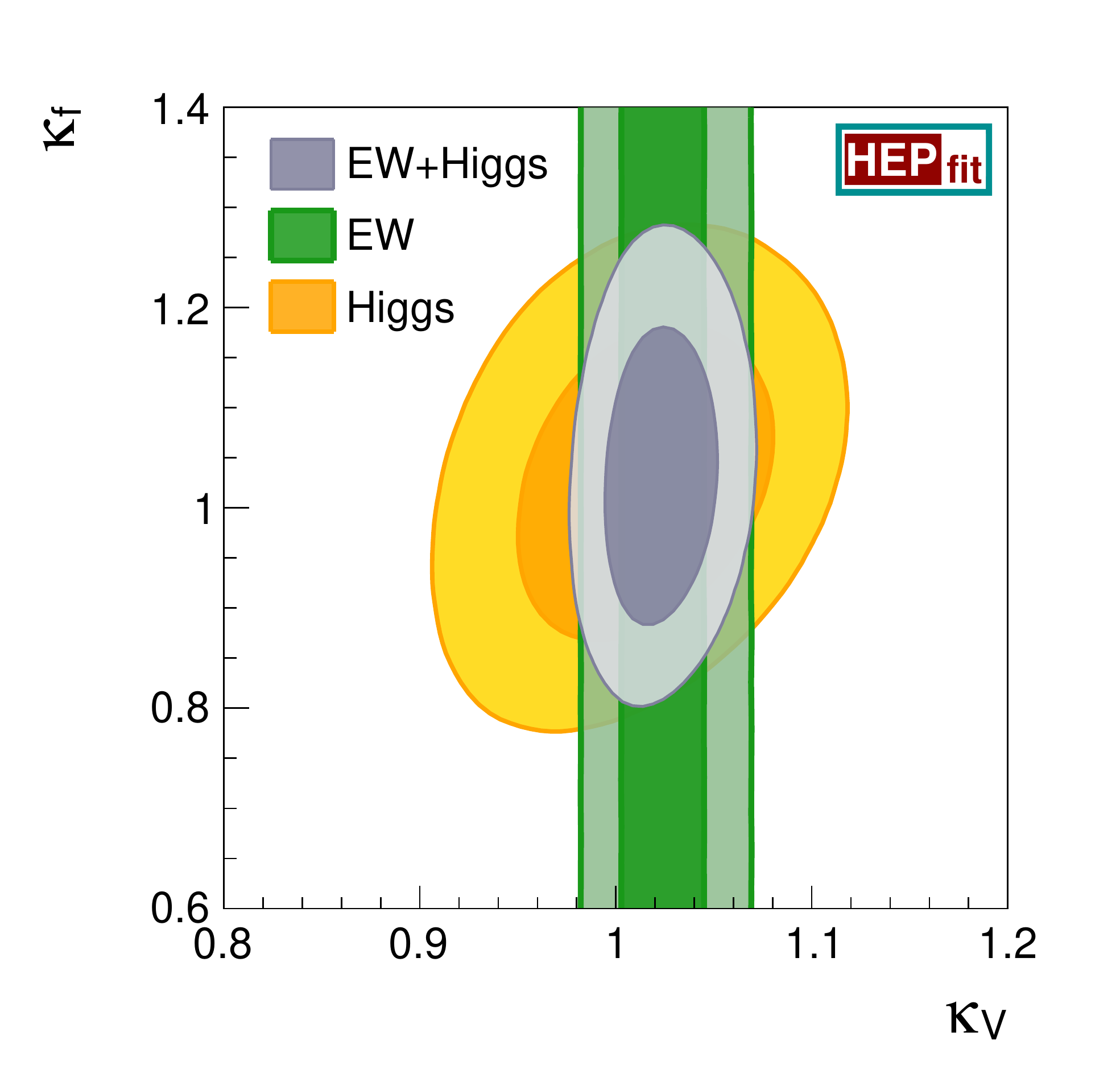} 
&
 \hspace{-0.3cm}\includegraphics[width=.317\textwidth]{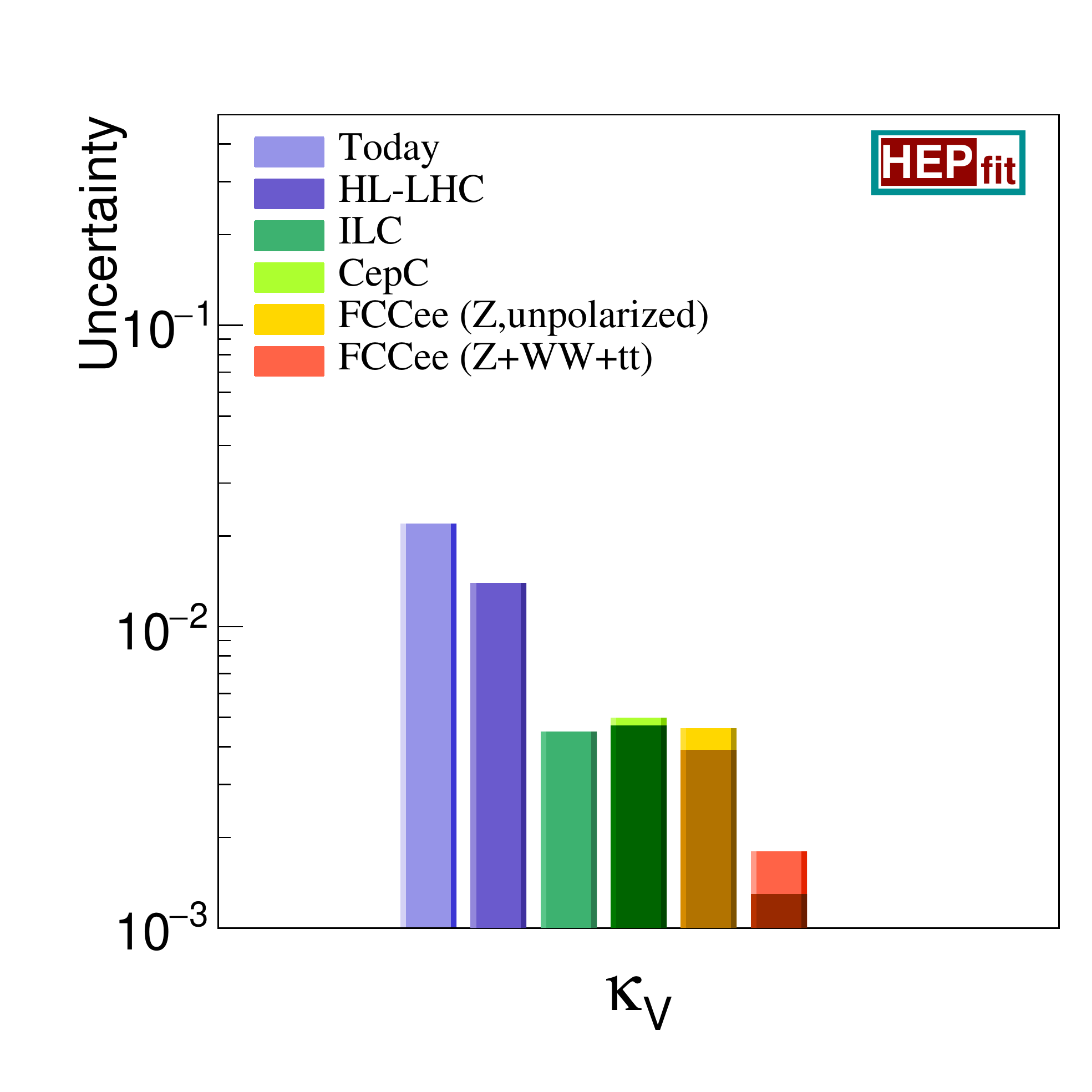}
  \end{tabular}
  \vspace{-0.65cm}
  \caption{(Left) 1D probability distribution for $\kappa_V$ derived from EWPD.  (Center)
  Comparison of the 68$\%$ and 95$\%$ probability contours for rescaled Higgs couplings to fermions ($\kappa_f$)
  and vector bosons ($\kappa_V$), from EWPO and Higgs signal strengths (see \cite{deBlas:2016ojx} for details). 
  (Right) Expected sensitivities to $\kappa_V$
    at future colliders. Different shades of the same colour correspond to
    results including or neglecting the future theoretical
    uncertainties. }
  \label{fig:HVV}
  \end{center}
\end{figure}

Next we study the EWPD constraints on NP models whose leading observable
effects appear in modifications of the Higgs couplings (see, e.g., Ref.~\cite{Contino:2010mh}). 
Assuming the new dynamics respects custodial symmetry, the deviations in the
Higgs to vector boson couplings can be parameterized by a single scale
factor $\kappa_V$ ($\kappa_V=1$ in the SM). This induces the leading effects
in EWPO, in the form of logarithmic contributions to the $S$ and $T$ parameters~\cite{Azatov:2012bz}. 
From the fit results in the left panel of Fig.~\ref{fig:HVV}, 
\begin{equation}
\kappa_V=1.02\pm 0.02,~\mbox{and}~\kappa_V\in [0.98,\, 1.07]~\mbox{at 95\% probability.}
\end{equation}
We also find a preference for $\kappa_V>1$, with 90$\%$ of probability. 
This imposes significant constraints on composite Higgs models, which generate values of 
$\kappa_V<1$, unless extra contributions to the oblique parameters are present. It is noteworthy
that, as can be seen in the central panel of Fig.~\ref{fig:HVV}, the EWPO constraints still dominate
the LHC run 1 bounds from Higgs signal strengths~\cite{deBlas:2016ojx}. 

Finally, we consider the general parametrization of NP effects using the SM effective field theory up to dimension 6. Assuming that the fields and symmetries of nature at energies below a given cutoff $\Lambda$ are those of the SM, the most general Lorentz and SM gauge invariant theory describing effects at those energies can be parametrized by
\begin{equation}
{\cal L}_{\mathrm{Eff}}={\cal L}_{\mathrm{SM}}+\sum_{d>4}\frac{1}{\Lambda^{d-4}}{\cal L}_d,~~\mbox{with}~~{\cal L}_d=\sum_i c_i {\cal O}_i,~~[{\cal O}_i]=d,
\end{equation}
where the invariant operators ${\cal O}_i$ are built solely using SM fields. The operator coefficients, $c_i$, encode all the information of the NP, and can be obtained upon matching with the UV completion~\cite{delAguila:2000rc}.
A complete set of non-redundant dimension 6 interactions includes a total of 59 operators. 
We use the basis presented in Ref.~\cite{Grzadkowski:2010es}, where only 10 operators contribute to EWPO. 
These include 2 bosonic interactions: 
${\cal O}_{\phi D}=\left|\phi^\dagger D^\mu \phi\right|^2$ and ${\cal O}_{\phi WB}=\left(\phi^\dagger \sigma_a \phi\right) W_{\mu\nu}^a B^{\mu\nu}$,
with $\sigma_a$ the Pauli matrices; 7 operators involving fermionic currents of the form
${\cal O}_{\phi \psi}^{(1)}=\left(\phi^\dagger \lrD^\mu \phi\right) \left(\overline{\psi}\gamma_\mu \psi\right)$ ($\psi=l,~e,~q,~u~,d$) and ${\cal O}_{\phi F}^{(3)}=\left(\phi^\dagger \sigma_a \lrD^\mu \phi\right) \left(\overline{F}\gamma_\mu \sigma_a F\right)$ ($F=l,~q$); and the four-lepton operator, ${\cal O}_{ll}=\left(\overline{l}\gamma_\mu l\right) \left(\overline{l}\gamma^\mu l\right)$.

\begin{figure}[h]
\begin{floatrow}
\capbtabbox{
\centering
{
\begin{tabular}{c c c }
 \toprule
&\multicolumn{2}{c}{95$\%$ prob. bound on $\frac{c_i}{\Lambda^2}$ [TeV$^{-2}$]} \\
Operator &1 op. at a time&Global \\
\cmrule
$\!\!{\cal O}_{\phi WB}\!\!$&$ [-0.009, 0.006]$ & --- \\
$\!\!{\cal O}_{\phi D}\!\!$&$ [-0.031, 0.006]$ & --- \\
$\!\!{\cal O}_{\phi l}^{(1)}\!\!$&$ [-0.006, 0.011]$ &$ [-0.013, 0.034]$ \\
$\!\!{\cal O}_{\phi l}^{(3)}\!\!$&$ [-0.012, 0.006]$ &$ [-0.065, 0.008]$ \\
$\!\!{\cal O}_{\phi e}^{(1)}\!\!$&$ [-0.017, 0.005]$ &$ [-0.028, 0.009]$ \\
$\!\!{\cal O}_{\phi q}^{(1)}\!\!$&$ [-0.025, 0.046]$ &$ [-0.099, 0.077]$ \\
$\!\!{\cal O}_{\phi q}^{(3)}\!\!$&$ [-0.011, 0.016]$ &$ [-0.179, 0.007]$ \\
$\!\!{\cal O}_{\phi u}^{(1)}\!\!$&$ [-0.065, 0.091]$ &$ [-0.230, 0.410]$ \\
$\!\!{\cal O}_{\phi d}^{(1)}\!\!$&$ [-0.159, 0.054]$ &$ [-1.11, -0.110]$\\
$\!\!{\cal O}_{ll}\!\!$&$ [-0.012, 0.020]$ &$ [-0.087, 0.026]$ \\ 
\bottomrule
 \end{tabular}
 }
 }{
\caption{95$\%$ probability limits on the dimension 6 operator coefficients entering in EWPD. See text for details.}
\label{tab:Dim6}
}\hspace{-0.6cm}
\ffigbox{
\centering
  \hspace{-0.1cm}\includegraphics[width=.45\textwidth]{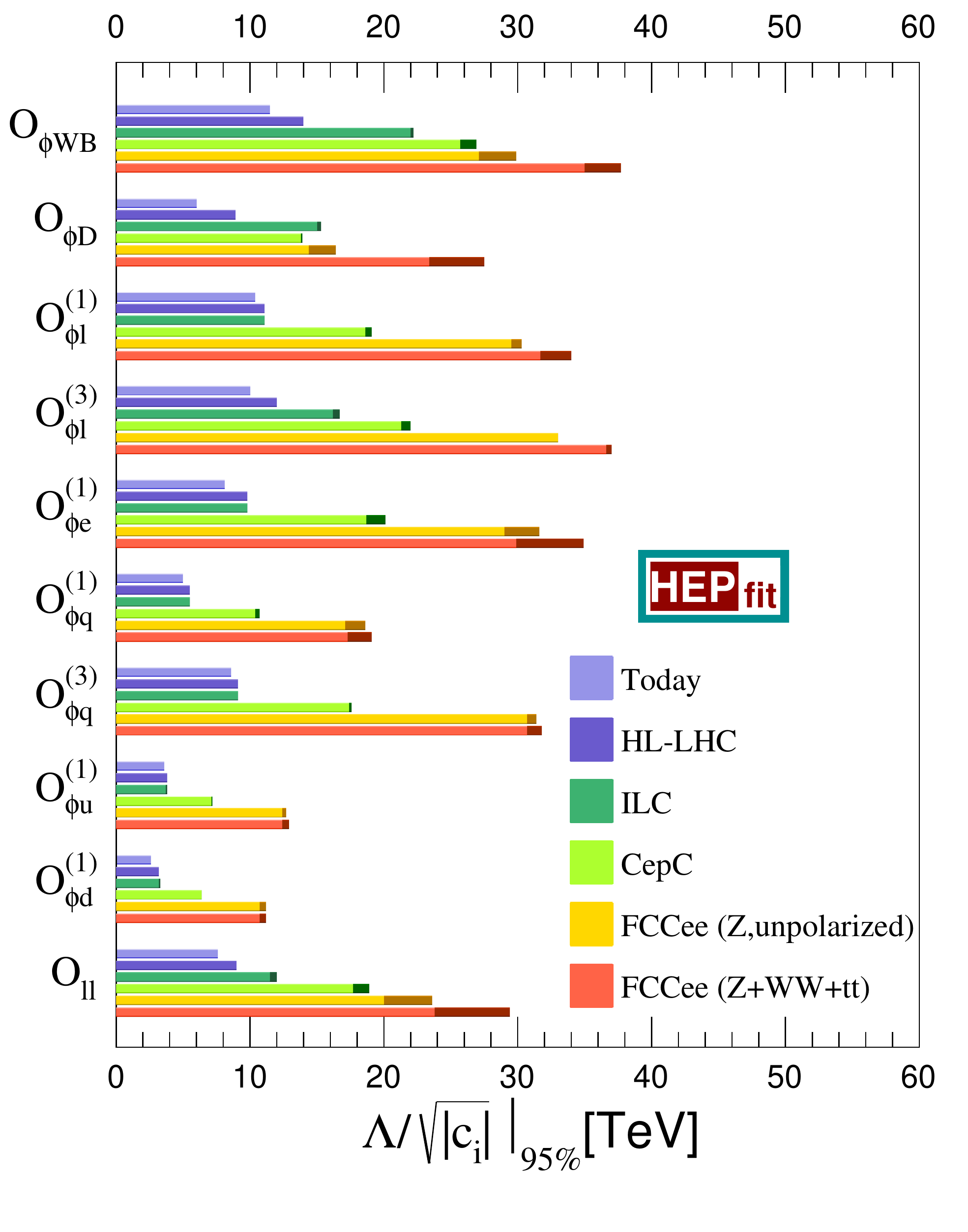}
  \vspace{-0.6cm}
}{
  \caption{Projected sensitivities to dimension 6 interactions at future colliders
  (1 operator at a time). Different shades of the same colour denote
    results including or neglecting future theory
    uncertainties.~\label{fig:futuredim6}}
}
\end{floatrow}
\end{figure}

After EWSB ${\cal O}_{\phi WB}$ and ${\cal O}_{\phi D}$ give rise to tree-level contributions to the $S$ and $T$ parameters, i.e. $\alpha_{\mathrm{em}} S=4\sin{\theta_W}\cos{\theta_W} c_{\phi WB}\mathrm{v}^2/\Lambda^2$ and $\alpha_{\mathrm{em}} T= -c_{\phi D} \mathrm{v}^2/(2\Lambda^2)$, while the ${\cal O}_{\phi \psi}^{(1,3)}$ induce corrections
to the neutral and charged current vertices. Finally, ${\cal O}_{ll}$ modifies the amplitude of muon decay, the process we use to extract the value of the Fermi constant $G_F$. This is one of the inputs of the SM and therefore this effect propagates to all EWPO. 
The results of the EWPD fit, assuming only one operator is generated by the NP, are given in the first column of Table~\ref{tab:Dim6}. Fig.~\ref{fig:futuredim6}, on the other hand, shows our preliminary estimates for the future projections. While for $|c_i| \sim 1$ current bounds could be interpreted as limits on the cut-off scale $\Lambda\gtrsim 3$-$12$ TeV at 95$\%$ probability, future experiments could probe values of $\Lambda\gtrsim 10$-$38$ TeV, depending on the interaction and the level of accuracy of future SM calculations.

While all of the 10 operators introduced above do enter in EWPO, the data only allows to constrain 8 independent combinations of the corresponding $c_i$. Using a field redefinition one can trade, e.g., ${\cal O}_{\phi WB}$ and ${\cal O}_{\phi D}$ for other 2 interactions that do not contribute to EWPO (but can be constrained using Higgs physics), and obtain a basis where the EWPD fit does not have any flat direction.
The results of a global fit to all the operators in such a basis are given in the second column in Table~\ref{tab:Dim6}. As can be seen, in this case the bounds are weaker with respect to the case of one operator at a time, which indicates the presence of large correlations. Only ${\cal O}_{\phi d}^{(1)}$ is not compatible with the SM at $95\%$ probability, 
with this deviation being caused by the $A_{\rm FB}^{0,b}$.
A more detailed discussion of this fit and other global analyses will be presented in a forthcoming publication.  (See also~\cite{Ciuchini:2013pca} for recent related work).

\section{Conclusions}

In this proceedings we have presented a brief summary of the sensitivities to NP from current EWPO and the future expected improvements at future $e^+ e^-$ colliders. While here we focused only on the limits from EWPD, some of the scenarios we discussed can be also constrained by other types of physics (e.g., Higgs observables). In many cases, however, the global bounds on NP are still dominated by EWPD~\cite{deBlas:2016ojx,deBlas:2014ula}, as can be seen, e.g., in Fig.~\ref{fig:HVV}. More generally, the EWPO provide complementary information to that from LHC searches, and are still as relevant as ever after the Higgs discovery. For further details and other analyses including the combination of EWPO and Higgs signal strengths see~\cite{deBlas:2016ojx}.



\vspace{0.cm}

\begin{thebibliography}{99}

\vspace{-0.07cm}
  
\bibitem{deBlas:2016ojx}
  J.~de Blas, M.~Ciuchini, E.~Franco, S.~Mishima, M.~Pierini, L.~Reina and L.~Silvestrini,
  JHEP {\bf 1612} (2016) 135
  [arXiv:1608.01509 [hep-ph]].
  
\vspace{-0.04cm}
  
\bibitem{Gomez-Ceballos:2013zzn}
  M.~Bicer {\it et al.} [TLEP Design Study Working Group Collaboration],
  JHEP {\bf 1401} (2014) 164
  [arXiv:1308.6176 [hep-ex]].
  
\vspace{-0.04cm}

\bibitem{Fujii:2015jha}
  K.~Fujii {\it et al.},
  arXiv:1506.05992 [hep-ex];
%
  T.~Barklow {\it et al.},
  arXiv:1506.07830 [hep-ex].

\vspace{-0.04cm}
  
\bibitem{CEPC}
  CEPC-SPPC Study Group,
  {\em CEPC-SPPC Preliminary Conceptual Design Report}, 2015.
  
\vspace{-0.04cm}
  
\bibitem{Peskin:1991sw}
  M.~E.~Peskin and T.~Takeuchi,
  Phys.\ Rev.\ D {\bf 46} (1992) 381.
  
\vspace{-0.04cm}
  
\bibitem{Contino:2010mh}
  R.~Contino {\it et al.},
  JHEP {\bf 1005} (2010) 089
  [arXiv:1002.1011 [hep-ph]].
  
\vspace{-0.04cm}
  
\bibitem{Azatov:2012bz}
  A.~Azatov, R.~Contino and J.~Galloway,
  JHEP {\bf 1204} (2012) 127
  [arXiv:1202.3415 [hep-ph]].
  
\vspace{-0.04cm}

\bibitem{delAguila:2000rc}
  F.~del Aguila, M.~Perez-Victoria and J.~Santiago,
  JHEP {\bf 0009} (2000) 011
  [hep-ph/0007316];
%
  F.~del Aguila, J.~de Blas and M.~Perez-Victoria,
  Phys.\ Rev.\ D {\bf 78} (2008) 013010
  [arXiv:0803.4008 [hep-ph]];
%
  JHEP {\bf 1009} (2010) 033
  [arXiv:1005.3998 [hep-ph]];
 %
  J.~de Blas {\it et al.},
  JHEP {\bf 1504} (2015) 078
  [arXiv:1412.8480 [hep-ph]].
  
\vspace{-0.04cm}

\bibitem{Grzadkowski:2010es}
  B.~Grzadkowski {\it et al.},
  JHEP {\bf 1010} (2010) 085
  [arXiv:1008.4884 [hep-ph]].
  
\vspace{-0.04cm}

\bibitem{Ciuchini:2013pca}
  M.~Ciuchini {\it et al.},
  JHEP {\bf 1308} (2013) 106
  [arXiv:1306.4644 [hep-ph]];
  F.~del Aguila and J.~de Blas,
  Fortsch.\ Phys.\  {\bf 59} (2011) 1036
  [arXiv:1105.6103 [hep-ph]];
%
  J.~de Blas,
  EPJ Web Conf.\  {\bf 60} (2013) 19008
  [arXiv:1307.6173 [hep-ph]].
%
  J.~de Blas, M.~Chala and J.~Santiago,
  Phys.\ Rev.\ D {\bf 88} (2013) 095011
  [arXiv:1307.5068 [hep-ph]];
%
  JHEP {\bf 1509} (2015) 189
  [arXiv:1507.00757 [hep-ph]];
  L.~Berthier and M.~Trott,
  JHEP {\bf 1602} (2016) 069
  [arXiv:1508.05060 [hep-ph]].
  
\vspace{-0.04cm}
  
\bibitem{deBlas:2014ula}
  J.~de Blas {\it et al.},
  Nucl.\ Part.\ Phys.\ Proc.\  {\bf 273-275} (2016) 834
  [arXiv:1410.4204 [hep-ph]];
%
  PoS EPS {\bf -HEP2015} (2015) 187.

\end{thebibliography}
\end{document}